\documentclass[aps,prb,showpacs,showkeys,twocolumn,amsmath,amssymb,reprint]{revtex4-1}

\usepackage{times}
\usepackage{color}
\usepackage{graphicx}
\usepackage{amsmath,amsbsy,amsfonts,mathrsfs}

\newcommand{\be}{\begin{equation}}
\newcommand{\ee}{\end{equation}}
\newcommand{\bea}{\begin{eqnarray}}
\newcommand{\eea}{\end{eqnarray}}
\newcommand{\bg}{\begin{figure}}
\newcommand{\eg}{\end{figure}}
\newcommand{\bi}{\begin{itemize}}
\newcommand{\ei}{\end{itemize}}

\usepackage{dcolumn}
\usepackage{textcomp}

\begin{document}
\bibliographystyle{apsrev}

\title[Chemical interaction, space-charge layer and molecule charging energy for metal oxide / organic interfaces]{Chemical interaction, space-charge layer and molecule charging energy for metal oxide / organic interfaces}

\author{\bf Jos\'e I. Mart\1nez}
\email{joseignacio.martinez@icmm.csic.es}
\address{Dept. Surfaces, Coatings and Molecular Astrophysics, Institute of Material Science of Madrid (ICMM-CSIC), ES-28049 Madrid (Spain)}

\author{\bf Fernando Flores}
\address{Dept. F\'{\i}sica Te\'orica de la Materia Condensada and Condensed Matter Physics Center (IFIMAC), Universidad Aut\'onoma de Madrid, ES-28049 Madrid (Spain)}

\author{\bf Jos\'e Ortega}
\address{Dept. F\'{\i}sica Te\'orica de la Materia Condensada and Condensed Matter Physics Center (IFIMAC), Universidad Aut\'onoma de Madrid, ES-28049 Madrid (Spain)}

\author{\bf Sylvie Rangan}
\address{Dept. Physics and Astronomy, and Laboratory for Surface Modification, Rutgers, The State University of New Jersey, Piscataway, NJ 08854-8019 (USA)}

\author{\bf Charles Ruggieri}
\address{Dept. Physics and Astronomy, and Laboratory for Surface Modification, Rutgers, The State University of New Jersey, Piscataway, NJ 08854-8019 (USA)}

\author{\bf Robert A. Bartynski}
\address{Dept. Physics and Astronomy, and Laboratory for Surface Modification, Rutgers, The State University of New Jersey, Piscataway, NJ 08854-8019 (USA)}

\date{\today}

\begin{abstract} 
Three driving forces control the energy level alignment between transition-metal oxides and organic materials: the chemical interaction between the two materials, the organic electronegativity and the possible space charge layer formed in the oxide. This is illustrated in this letter by analyzing experimentally and theoretically a paradigmatic case, the TiO$_2$(110) / TCNQ interface: due to the chemical interaction between the two materials, the organic electron affinity level is located below the Fermi energy of the $n$-doped TiO$_2$. Then, one electron is transferred from the oxide to this level and a space charge layer is developed in the oxide inducing an important increase in the interface dipole and in the oxide work function. 
\end{abstract}



\maketitle


Hybrid materials that contain interfaces between transition metal oxides and organic species exhibit very promising properties for applications in devices like solar cells, light emitting diodes, fuel cells or thin films transistors. In particular, the easy injection of charge between the oxide and the organic, which depends critically on the barriers formed at the interface between the two materials, plays a very important role in the good efficiency of those devices.~\cite{ForrestCR1997,SessoloAM2011,OreganN1991,KabraAM2010,MeyerAM2012,LiCM2014}

A large amount of work has been undertaken in an effort to understanding the energy level alignment between different metal/organic and organic/organic interfaces.~\cite{FloresPCCP2009,CahenMT2005,KochJPCM2008} In contrast, very few studies have analyzed the energy level alignment at the interface between transition metal oxides and organic semiconductors. In a recent work, Greiner {\it et al.}~\cite{GreinerNM2011} analyzed a variety of non-reactive oxide/organic interfaces and concluded that the energy level alignment is determined by one driving force: the electron chemical potential equilibration between the oxide Fermi level and the organic ionization energy.  On the other hand, Xu {\it et al.}~\cite{XuPRL2013} have conclusively shown that a second driving force is the oxide doping and the concomitant formation of a space charge layer upon the interaction with the organic material; for strongly $n$-doped oxides, like ZnO or TiO$_2$, this mechanism is particularly important when the organic affinity level is located below the oxide Fermi level, as is the case of F4TCNQ physisorbed on a H-saturated ZnO(000$\bar{1}$) surface.~\cite{XuPRL2013} In this letter we show that a third driving force, the chemical interaction between the oxide surface and the organic material, favors the charge transfer between the two materials and thus plays a fundamental role for the energy level alignment in oxide/organic interfaces.

\begin{figure}[t]
\centerline{\includegraphics[width=\columnwidth]{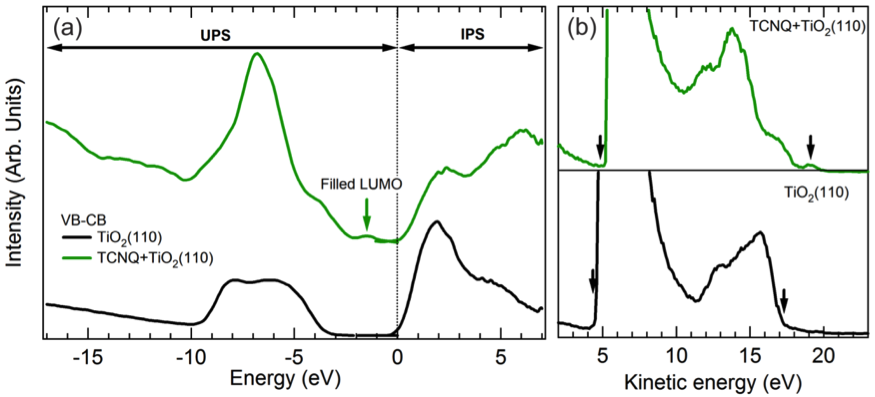}}
\smallskip \caption{(color online) (a) Valence and conduction band spectra measured using UPS and IPS, respectively, of the clean 
TiO$_2$(110) surface and of the same surface saturated with TCNQ. The zero of energy is chosen as the position of the Fermi level.
(b) Secondary electron cutoff determination using the full width of the emitted photoelectrons.
\label{Fig1}} 
\end{figure}

In order to understand how the chemical interaction determines the oxide/organic energy level alignment, we have analyzed in this work the particular case of the TiO$_2$ / TCNQ interface; TiO$_2$ is one of the most extensively studied substrates for organic devices~\cite{SanchezNS2013, SanchezJPCC2014}, while TCNQ (tetracyanoquinodimethane) is an organic molecule frequently used due to its very electronegative properties~\cite{FerrarisJACS1973,AlvesNM2008}, showing a strong chemical interaction with the substrate. 
We can expect that an important chemical interaction and charge transfer should appear between the two materials, which should affect the interface barrier formation as well as the creation 
of a metal oxide space charge. In the following we shall prove experimentally and theoretically that this charge transfer is extreme with one electron being transferred from the oxide 
to the LUMO$_{\textrm{TCNQ}}$ level. Our results are compatible with an important increase in the interface barrier and in the metal oxide work-function. We shall also show that the 
chemical interaction between the oxide and this very electronegative organic material plays an important role in the creation of that charge transfer and in the formation of an oxide space charge. 
All these effects should be considered as an important ingredient in the design of devices with very electronegative organic materials.


\begin{figure}[t]
\centerline{\includegraphics[width=\columnwidth]{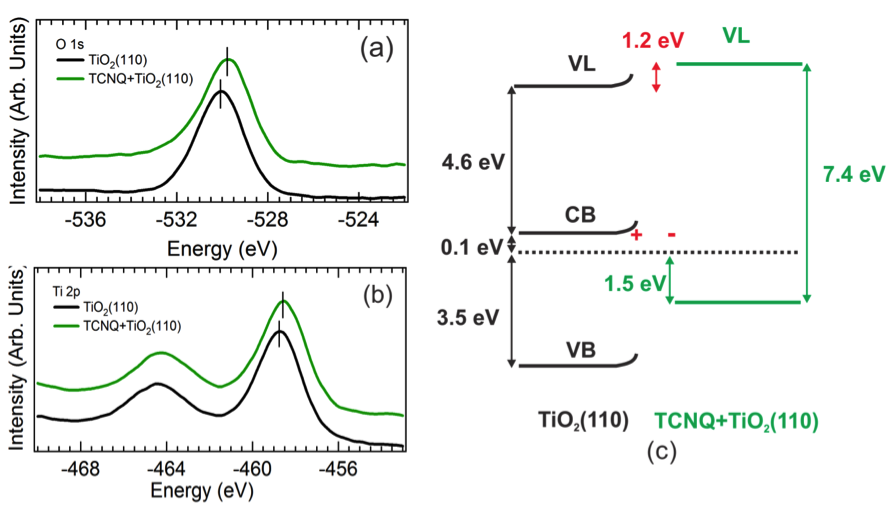}}
\smallskip \caption{(color online) (a) O 1s and (b) Ti 2p core levels of the substrate before and after TCNQ adsorption. A clear peak displacement is observed for both core levels and attributed to upward band bending. (c) Energy diagram obtained from the data shown in Figures~\ref{Fig1} and~\ref{Fig2}.
\label{Fig2}} 
\end{figure}

A TiO$_2$(110) single crystal surface prepared in ultra-high vacuum (see Supplemental Material (SM) for further details) and exposed to TCNQ was found to saturate at room temperature to a coverage referred to as monolayer coverage (ML) in the following work. The valence and conduction band spectra of the clean and TCNQ-exposed TiO$_2$, measured, respectively, using UPS and IPS are displayed in Figure~\ref{Fig1}(a). On this figure, the zero of energy is defined as the Fermi level, so that the occupied states are characterized with a negative energy and the unoccupied state with a positive energy. The valence band of the pristine TiO$_2$(110) surface originates mainly from O 2p states, while the conduction band is composed of Ti 3d states. A linear fit of the sharp band edges to the background of the spectra indicates a valence band maximum at -3.5 eV and a conduction band minimum at 0.1 eV, resulting in a 3.6 eV gap for TiO$_2$. The position of the Fermi level, only 0.1 eV below the conduction band, is indicative of the strong n-doped nature of the TiO$_2$ crystal. Upon TCNQ adsorption, molecular states are appearing both in occupied and unoccupied states as seen in Figure~\ref{Fig1}(a). These molecular states cannot be interpreted in terms of the molecular signature of an intact TCNQ molecule. In contrast, the valence band (VB) and the conduction band (CB) spectra of a TCNQ multilayer grown at 230K on a metal substrate (shown in SM) can be directly compared to the DOS calculated for a TCNQ molecule. This indicates that TCNQ is strongly affected by the presence of the TiO$_2$ surface. Particularly important for this study, the first occupied molecular states are found within the gap of TiO$_2$, 1.5 eV below the measured Fermi level, as indicated by the arrow in Figure~\ref{Fig1}(a). In the unoccupied states, broad molecular features, superposed upon the contribution of the strong Ti 3d state of the TiO$_2$ substrate CB, prevent a clear determination of the unoccupied frontier molecular states.

The position of the vacuum level of the system has also been measured for the clean and TCNQ exposed TiO$_2$(110) surface, using the position of the secondary electron cutoff (SECO) of the total spectra of emitted photoelectrons shown in Figure~\ref{Fig1}(b). An energy separation of 13 eV (delimited by the arrows) is measured between the VB edge and the SECO of the clean TiO$_2$(110) surface. With a photon excitation energy of 21.2 eV and a measured gap of 3.6 eV, the electron affinity for the TiO$_2$(110) surface is found to be 4.6 eV. For the TCNQ-saturated TiO$_2$ surface, the distance between the first occupied molecular states and the SECO (delimited by the arrows) is measured to be 13.8 eV, resulting in a distance of 7.4 eV between that first occupied molecular state and the VL (vacuum level) of the molecule (see Figure~\ref{Fig2}(c)).

Figure~\ref{Fig2} shows XPS spectra measured on the clean and on the subsequently saturated TiO$_2$(110) surface. In Figure S1 (see SM) large scale survey scans indicate that, as expected, only C 1s and N 1s core levels are added to the initial Ti 2p and O 1s core levels belonging to the surface (see Figure S1 in SM). The molecular coverage can be evaluated by comparing the relative ratio of C 1s and Ti 2p core levels to no more than a monolayer. Upon TCNQ adsorption, a noticable shift of the TiO$_2$ surface core levels is observed as shown in Figure~\ref{Fig2}(a) and Figure~\ref{Fig2}(b). Both the O 1s and Ti 2p core level spectra are found shifted toward the Fermi level by about 0.2 eV after TCNQ adsorption. Such behavior is interpreted as an upward band bending at the surface of the TiO$_2$ substrate, due to charge reorganization at the TiO$_2$ / TCNQ interface. This band bending and the interface VL shift shown in Figure~\ref{Fig2}(c) are indicative of a strong electron charge transfer from the oxide to the molecule. This is analyzed theoretically in the following.


We have analyzed theoretically the TiO$_2$ / TCNQ interface by means of a local-orbital Density Functional Theory (DFT) approach (see section 2 in SM). In a first step, we have considered the T=0K case and analyzed the interface geometry neglecting van der Waals interactions due to the strong covalent bond between the molecule and the oxide. In a second step, the electronic properties of the interface, including the TiO$_2$ / TCNQ level alignment and charge transfer, are calculated introducing appropriate corrections in the DFT calculation~\cite{MartinezJCP2013,BeltranJPCC2013}.

\begin{figure}[h]
\centerline{\includegraphics[width=\columnwidth]{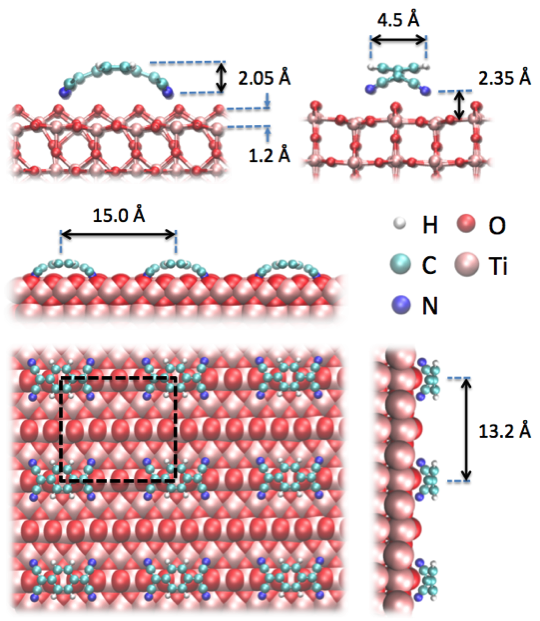}}
\smallskip \caption{(color online) Surface geometry for the TiO$_2$(110) / TCNQ system.
\label{Fig3}} 
\end{figure}

Lattice vectors and unit cell for the periodic DFT calculation are shown in Figure~\ref{Fig3} (\textbf{a}=13.2~\AA~and \textbf{b}=15.0~\AA). The TiO$_2$(110) surface is represented by a slab with 5 layers and the TCNQ adlayer is placed on one side of the slab; thus, there are in total 360 atoms per unit cell including the TCNQ molecule. The Brillouin zone (BZ) has been sampled by means of a [2$\times$4$\times$1] Monkhorst-Pack grid~\cite{MP}, guaranteeing a full convergence in energy and electronic density. In our calculations for the interface geometry, we have started with a perfectly flat TCNQ molecule and have applied a dynamical relaxation procedure to obtain the most stable chemisorbed state. Our DFT calculations show that the TCNQ molecule forms strong covalent bond with the TiO$_2$(110) surface. After several initial positions of the molecule on the oxide unit cell, we have obtained the relaxed interface geometry shown in Figure 3: TCNQ is deformed by its interaction with the oxide, with the N atoms strongly bonded to O and Ti; while the mean distance from the central part of the molecule to the O--first layer is 3.20~\AA, the N--O and N--Ti distances are around 2.8~\AA~and 2.87~\AA, respectively. The unit cell size defining the molecule-molecule distance has been fixed assuming a good matching between the oxide and the adsorbed TCNQ-structure; we stress that these distances are similar to the ones found in other TCNQ-interfaces~\cite{MartinezPSSB2011,Torrente2008,HuiJuan}.

For organic materials, it is well-known that in standard DFT calculations the Kohn-Sham energy levels do not properly describe the electronic energy levels of the system so that transport gaps are usually too small~\cite{FloresPCCP2009,MartinezJCP2013,BeltranJPCC2013}. For example, the experimental gap between the ionization and the affinity levels of the gas-phase TCNQ molecule is around 5.3 eV, while the energy gap between the Kohn-Sham HOMO and LUMO levels in LDA (or in GGA) calculations, E$_{\textrm{g}}^{\textrm{LDA}}$, is 1.65 eV~\cite{MilianCPL2004,KanaiAPA2009,MedjanikPRB2010}. This problem is related to the molecule self-interaction energy as described by the molecule charging energy~\cite{FloresPCCP2009}, U, that for the gas phase molecule is: U$_{\textrm{mol}}$=3.65 eV.

In the case of the organic-oxide interface, electron correlation effects (dynamical polarization) reduce the charging energy of the molecule at the interface from U$^{\textrm{mol}}$ to U (and its energy gap E$_{\textrm{g}}$ from E$_{\textrm{g}}^{\textrm{LDA}}$+U$^{\textrm{mol}}$ to E$_{\textrm{g}}^{\textrm{LDA}}$+U)~\cite{AbadOE2010}. Following previous works~\cite{FloresPCCP2009,BeltranPCCP2014,AbadJCP2011}, we can account for those effects in a practical and simplified way by introducing the following operator in the DFT calculation:
\be \label{eq1}
O_{\alpha}^{\textrm{scissor}}=\sum_{\mu\nu}\{(\epsilon+\frac{U}{2})|\mu_i\rangle\langle\mu_i|+(\epsilon-\frac{U}{2})|\nu_i\rangle\langle\nu_i|\},
\ee $|\mu_i\rangle$ and $|\nu_i\rangle$ being the empty (occupied) orbitals of the isolated molecule (with the actual geometry of the molecule on the surface). Using this operator, we can fix with U the initial value of the HOMO-LUMO gap (E$_{\textrm{g}}^{\textrm{LDA}}$+U), and with $\epsilon$ its relative position with respect to the oxide conduction band edge, which is not well described in conventional LDA calculations either. This is reminiscent of the ``shift and stretch'' procedure~\cite{SegevPRB2006} used to correct the DFT-DOS calculations. Although in our calculations the oxide energy gap is 3.0 eV instead of the experimental value of 3.6 eV, we define an appropriate initial alignment between the conduction band edge and the organic LUMO level, since we are interested in the possible charge transfer from the conduction band of the n-doped oxide to the organic LUMO level.

Due to dynamical polarization effects, we reduce U$^{\textrm{mol}}$=3.65 eV to U=1.95 eV, and take E$_{\textrm{g}}$=3.6 eV for an isolated molecule adsorbed on TiO$_2$. This energy gap is similar to, but a little larger than, the one found for the Au / TCNQ interface~\cite{MartinezPSSB2011} due to a smaller screening of the oxide in comparison with that of the metal surface. As discussed below, in our LDA calculations for the {\it full TCNQ-monolayer}, we find U$^{\textrm{0}}$, the charging energy that incorporates the interaction between molecules, to be 2.25 eV in good agreement with the value of U=1.95 eV taken here for an isolated molecule on the surface~\cite{MartinezPSSB2011}. On the other hand, $\epsilon$ is taken to yield the initial LUMO level (before the contact with the oxide is established) at 5.0 eV from vacuum, a value which is close to the one deduced from a TCNQ multilayer on Cu (4.4 eV with a broadening of 0.6 eV, see figure S2) if the energy gap is reduced symmetrically around the mid-gap to 3.6 eV, and chosen
to yield a better agreement with experiment (Figure~\ref{Fig2})~\cite{epsilon}.

\begin{figure}[t]
\centerline{\includegraphics[width=\columnwidth]{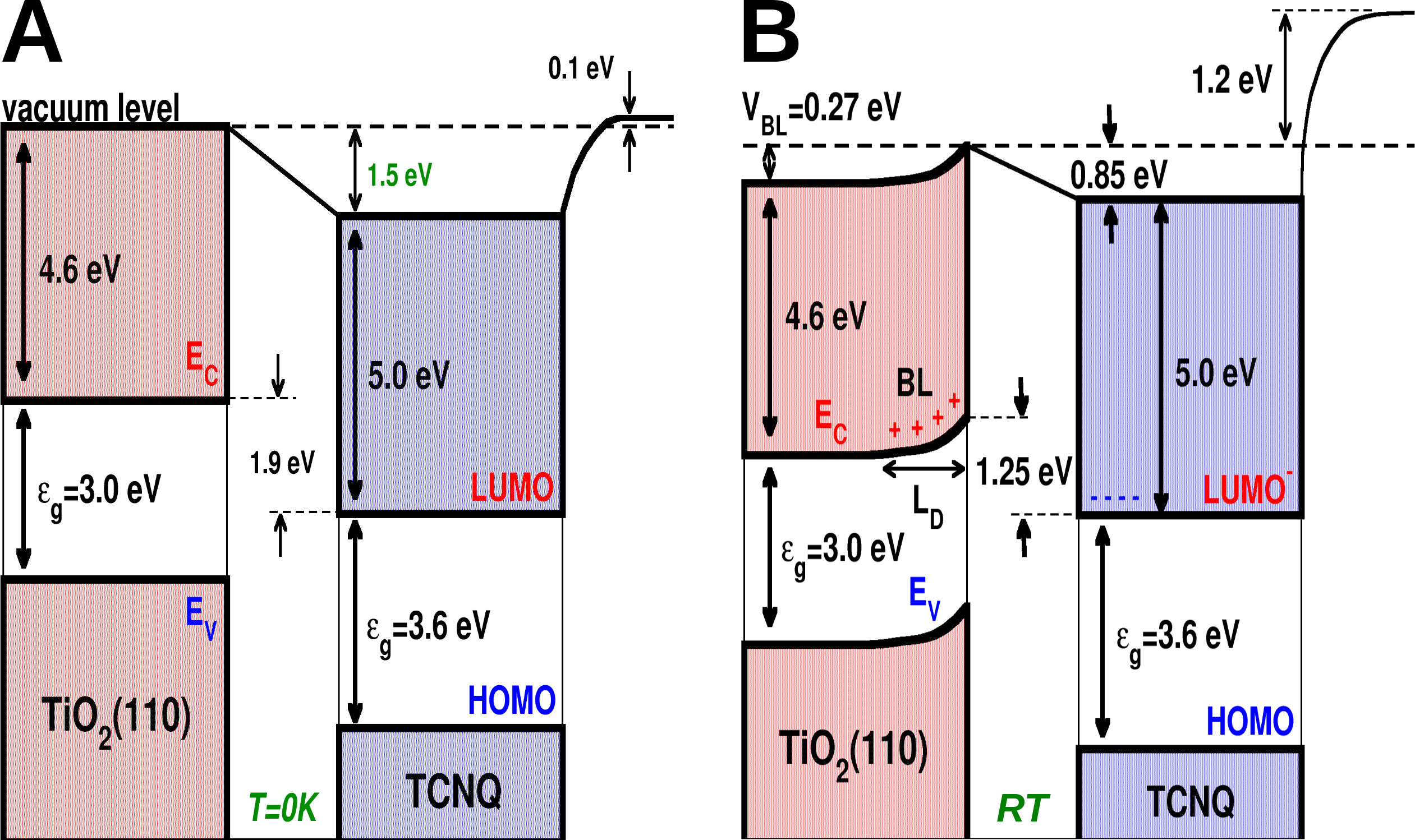}}
\smallskip \caption{(color online) (a) TiO$_2$ / TCNQ interface alignment for T=0K; (b) Idem for $RT$, showing the oxide band-bending.
\label{Fig4}} 
\end{figure}

Figure~\ref{Fig4}(a) shows the TiO$_2$ / TCNQ level alignment for T=0K as obtained in these calculations. The initial LUMO level is located 0.4 eV below the conduction band edge, E$_\textrm{C}$, but its final position is 1.9 eV below E$_\textrm{C}$. This LUMO level shift is due to two effects: (a) a the strong oxide-molecule hybridization, inducing a shift of 1.6 eV to deeper binding energies; and (b) an electrostatic dipole of 0.10 eV pushing the LUMO level upwards due to charge rearrangement upon hybridization; this dipole is partially due to a distortion of the molecule, V$^\textrm{mol}$ (eV$^\textrm{mol}$= -0.50 eV) and to charge transfer, V$^\textrm{charge}$ (eV$^\textrm{charge}$=0.60 eV). The result of the hybridization and the total electrostatic dipole is a downward displacement of the empty LUMO level with respect to its initial position of 1.50 eV, yielding a LUMO that is 1.9 eV below the conduction band.

This energy level alignment suggests that, at room temperature and for E$_\textrm{C}$-E$_\textrm{F}$=0.1 eV (n-doped TiO$_2$), there should be a strong thermally excited charge transfer from the oxide to the LUMO$_{\textrm{TCNQ}}$ (called LUMO'$_{\textrm{TCNQ}}$). This transfer of charge should create an important electrostatic potential at the interface with two contributions: a surface potential, V$_\textrm{BL}$, due to the space charge layer in the oxide extending a distance of a Debye length, L$_\textrm{D}$, into the crystal (see Figure~\ref{Fig4}b), and an interface potential due to the negative charge in the LUMO'$_{\textrm{TCNQ}}$ level (Q in electron units) and the opposite positive charge in the oxide (-Q). 
The space charge (or boundary layer BL) potential, V(z), can be analyzed classically with Poisson's equation~\cite{Frankl1967}:
\be \label{eq2}
L_\textrm{D}^{2}\frac{d^{2}(eV/k_{\textrm{B}}T)}{dz^{2}}=-[e^{-eV/k_{\textrm{B}}T}-1],
\ee where L$_{\textrm{D}}^{2}=\epsilon k_{B}T / 4\pi en$; $n$ being the electron charge density for the n-doped oxide material. In our calculations n$\approx$10$^{19}$ cm$^{-3}$, $\epsilon\approx$120 for the oxide (110) crystallographic direction, and L$_{\textrm{D}}\approx$50 \AA. On the other hand, the surface potential, V$_{\textrm{BL}}$, can be approximated by V$_{\textrm{BL}}=4\pi QeL_{\textrm{D}}/\epsilon A$, $A$ being the surface area per TCNQ  molecule (around 200 \AA$^{2}$), so that we have: eV$_{\textrm{BL}}/Q\approx$0.27 eV; this value is compatible with our measurements for the shift of the Ti core levels near the surface taking Q=1 (see Figure~\ref{Fig2}). As the Debye length is around 50 \AA, photoemission experiments sample only a small fraction of the boundary layer and the effect of the space charge on the organic / oxide alignment is to displace both the oxide surface layers and the organic levels to positive energies.

On the other hand, due to the charge transfer, Q, between the oxide and the LUMO'$_{\textrm{TCNQ}}$ level, there should appear a shift in the molecular levels given by eV$^{\textrm{S}}$=U$^\textrm{0}$Q. Due to this effect, the TCNQ levels are shifted upwards, for Q=1, by U$^\textrm{0}$, which we find to be around 2.25 eV: this quantity is calculated as the derivative of eV$^\textrm{charge}$ with respect to the charge transfer (in electron units), quantities which are obtained by introducing a fictitious shift, $\Delta$, to the TCNQ molecular levels (see figure S3 in SM).

We analyze how that induced potential, V$^\textrm{S}$, modifies the organic / oxide alignment by making use of the previous calculations for the case T=0K: the idea is to introduce in this approach an external shift, $\Delta_\textrm{0}$, to the TCNQ-levels which simulates the effect of that induced potential, V$^\textrm{S}$. This implies taking $\Delta_\textrm{0}$=2.25 eV for the TCNQ levels, and recalculating with this shift the new organic / oxide realignment~\cite{spin}. Figure~\ref{Fig4}b shows our results for the organic / oxide alignment after introducing this $\Delta_\textrm{0}$-shift. The first point to realize is that, in this case, the LUMO'$_{\textrm{TCNQ}}$ level is still 1.25 eV below the conduction band edge, to be compared with the experimental value of 1.8 eV (1.6+0.2 eV, see Figure~\ref{Fig2}c); this indicates that the charge transfer to the LUMO level is one electron, as mentioned above. The 1.25 eV value for (E$_\textrm{C}$-LUMO) is the result of a strong chemical shift of 2.05 eV and of a new induced interface dipole of 1.05 eV, associated with the electron charge transfer from the molecule to the oxide, which tends to oppose the $\Delta_\textrm{0}$-displacement. Notice that these effects change the oxide work-function by 1.20 eV (as yielded by the molecule charging energy, $\Delta_\textrm{0}$, and the induced interface dipole of 1.05 eV); including the boundary layer potential of 0.27 eV, we find a work-function change of 1.47 eV, in reasonable agreement with the experimental evidence of 1.2 eV.

We finally mention that a second TCNQ layer (or a multilayer) would feel an important realignment with respect to the oxide because the chemical shift of 2.05 eV, associated with the interaction between the oxide and the TCNQ first layer, should disappear; at the same time, we can expect an increase of the oxide energy gap to around 5.1 eV which would also shift the  LUMO'$_{\textrm{TCNQ}}$ level by 0.75 eV (1/2 of the change in the energy gap) to higher energies. These two effects should shift the  LUMO'$_{\textrm{TCNQ}}$ level from 1.25 eV below E$_\textrm{C}$ to 1.55 eV above, indicating that only the first TCNQ layer would develop a strong accumulation of charge.

In conclusion, we have shown that there is an important charge transfer between TiO$_2$ and the TCNQ monolayer, with one electron filling the LUMO level of the organic molecule. This is strongly suggested by the experimental evidence showing that, upon the deposition of a TCNQ monolayer on TiO$_2$, a space charge in the oxide is formed and that an important increase in the oxide work-function appears. Our theoretical analysis, based on a combination of a DFT approach and a calculation of the space charge layer potential, supports this interpretation and shows the important role that the oxide / organic interface chemistry, as well as their electron chemical equilibration and the oxide space charge layer, have in the barrier formation. Our results are tantamount to the formation of an electron accumulation in the first organic layer; although this strong accumulation of charge can be expected to disappear for successive layers, this effect should be considered as an important ingredient for tuning devices having those components.

SR, CR, and RAB acknowledge support from the National Science Foundation under Grant No NSF-CHE 1213727. JO and FF acknowledge support from Spanish MICIIN under contract FIS2010-16046.  JIM acknowledges funding from the CSIC-JAEDOC Fellowship Program, cofunded by the European Social Fund.

\bibliography{bib} 

\begin{thebibliography}{31}
\bibitem[{eps()}]{epsilon}
\bibinfo{note}{Fitting the initial LUMO level to 4.4 eV, at a position slightly
  above E$_C$, would shift upwards the final LUMO or LUMO' levels, see
  Figure~\ref{Fig4}, by around 0.3 eV.}
\end{thebibliography}

\end{document}